\documentclass[aps,twocolumn,showpacs,prl,floatfix]{revtex4}
\usepackage{graphicx}
\usepackage{epsf}
\usepackage{wrapfig}
\usepackage{epsfig}
\usepackage{sublabel}
\usepackage{epsfig}
\def\Vec#1{\mbox{\boldmath $#1$}}

\def\beq{\begin{equation}}
\def\eeq{\end{equation}}

\def\beqy{\begin{eqnarray}}
\def\eeqy{\end{eqnarray}}

\begin{document}
\vskip 2mm \date{\today}\vskip 2mm
\title{Proton-Proton and Proton-Neutron
Correlations in Medium-Weight Nuclei: Role of the
Tensor Force  within a Many-Body Cluster Expansion}
\author{M. Alvioli}
\author{C. Ciofi degli Atti}\affiliation{Department of Physics, University of Perugia and
      Istituto Nazionale di Fisica Nucleare, Sezione di Perugia,
      Via A. Pascoli, I-06123, Italy}
\author{H. Morita}\affiliation{Sapporo Gakuin University, Bunkyo-dai 11, Ebetsu 069,
  Hokkaido, Japan}

\begin{abstract}
A detailed analysis of the effect of tensor correlations on one- and two-body
densities and momentum distributions of complex nuclei is presented within a linked
cluster expansion providing reliable results for the ground state properties of nuclei
calculated with realistic interactions.
\end{abstract}
\maketitle Obtaining information on short range nucleon-nucleon
correlations (SRC) in nuclei is a primary goal of modern nuclear
physics. The interest in  SRC  stems not only from the necessity
to firmly establish the limits of validity of the standard model
of nuclei but also from the strong impact that the knowledge of
the short range structure of ordinary nuclei would have on other
fields of physics, like e.g.  nuclear physics of the stars and
astrophysics. As a matter of fact, when the distance between the
center-of-mass (CM) of two nucleons is about 1 $fm$, the local
density of of such a  pair  is several times larger than that of
the central density in nuclei  and comparable to that expected in
neutron stars; short range correlated NN pairs represent therefore
a form of cold dense nuclear matter that can be approached and
studied in the laboratory. Such a study, in particular the isospin
dependence of SRC,  would  help to answer several crucial
questions on the formation and the structure of neutron stars. As
a matter of fact,  in spite of the small probability of
neutron-neutron ($nn$) correlations a small concentration of
protons inside neutron stars is
\begin{figure}[!hbp]
\centerline{(a)\hspace{0.3cm}\epsfysize=0.42cm\epsfbox{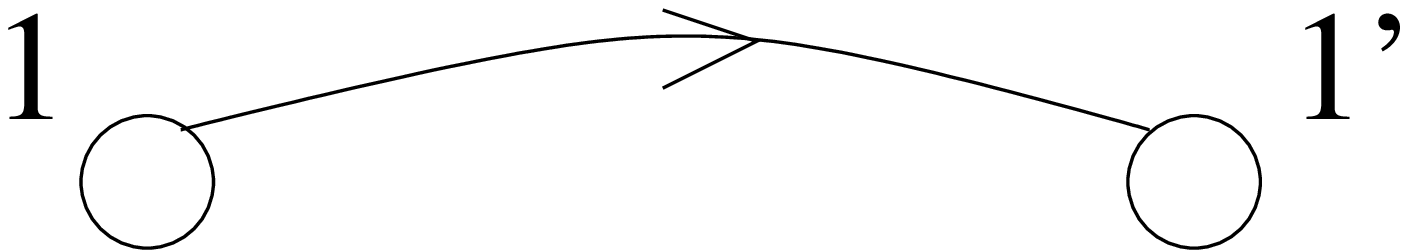}}
\vskip 0.2cm
 \centerline{
  \epsfysize=1.2cm\epsfbox{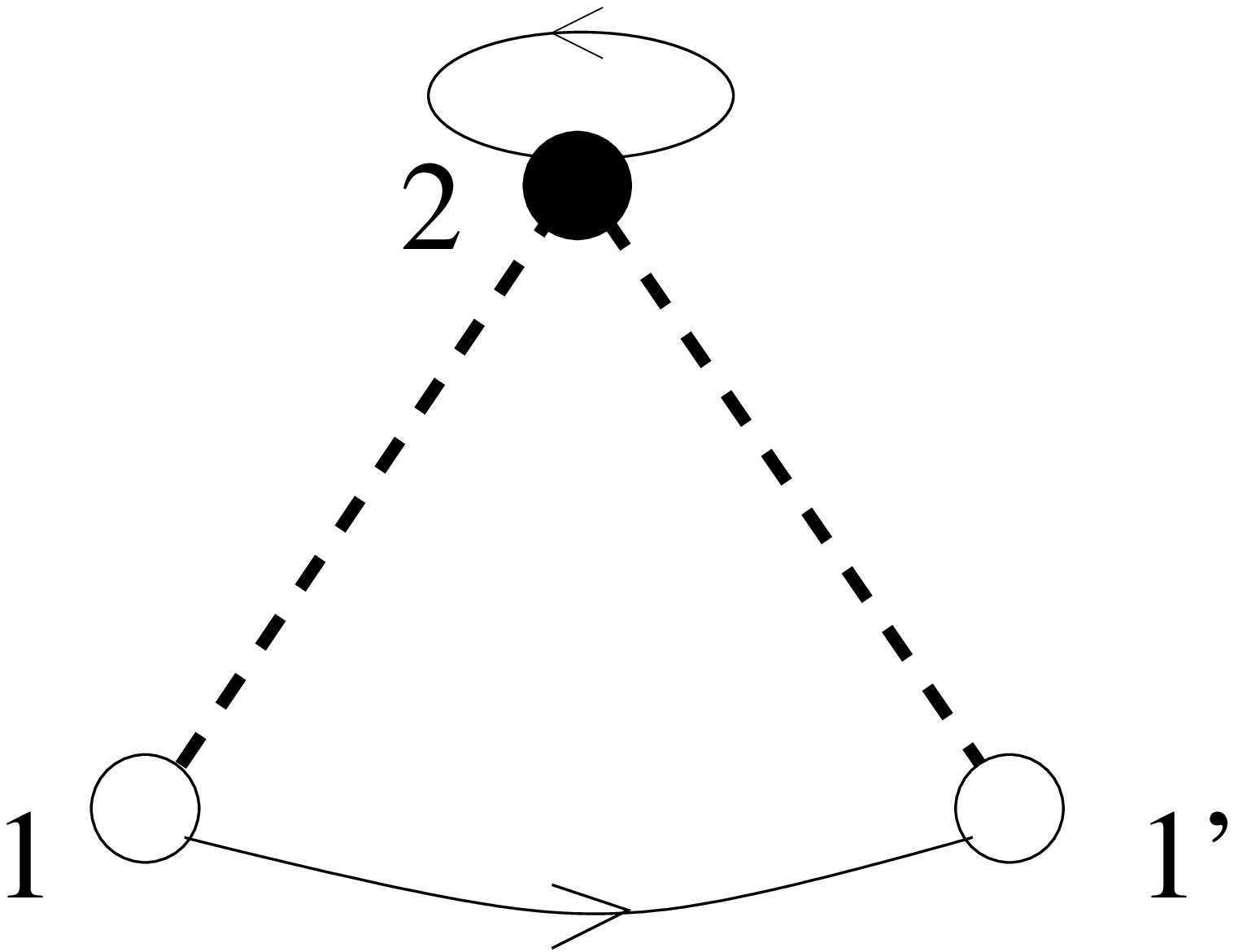}
  \epsfysize=1.2cm\epsfbox{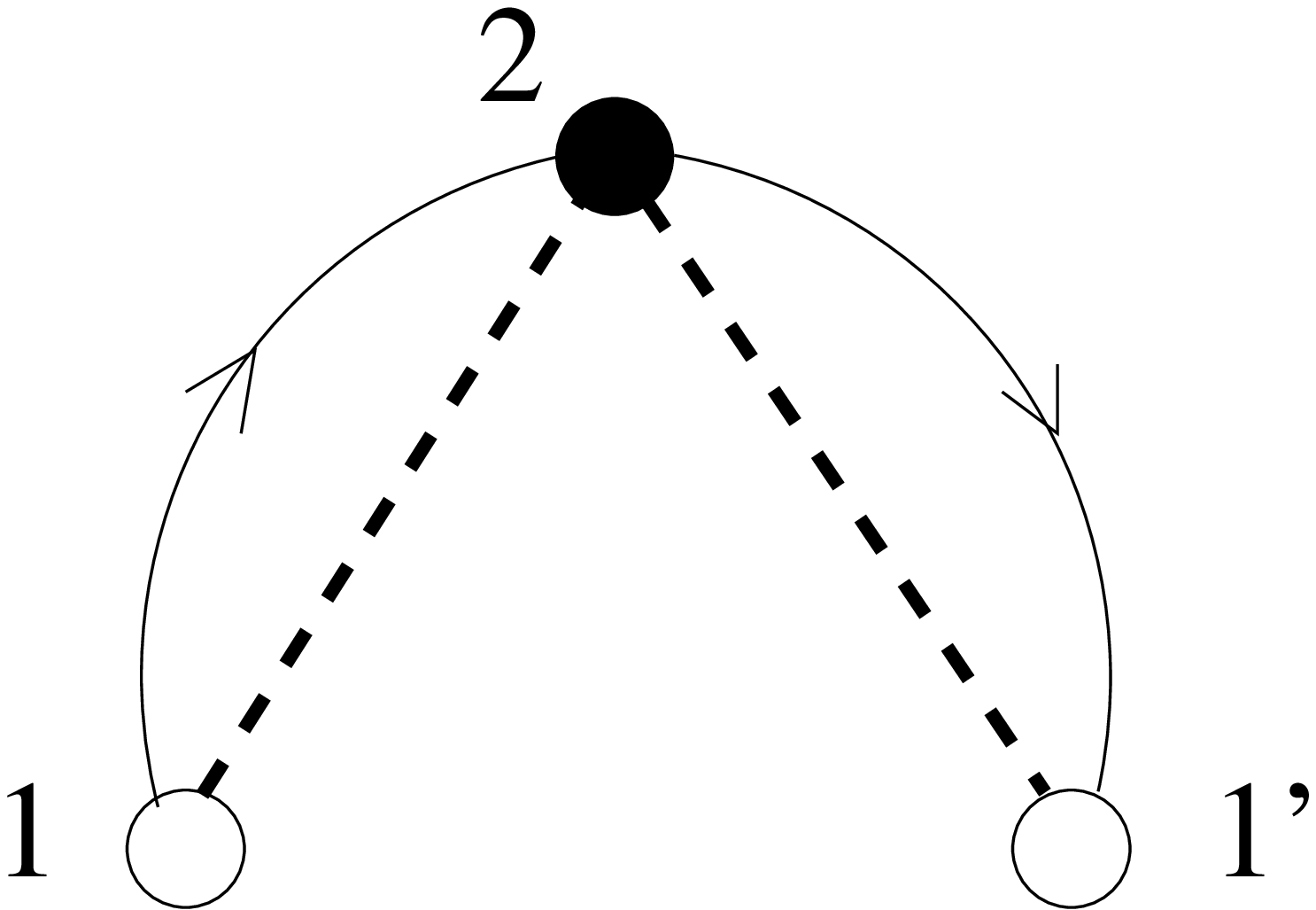}
  \epsfysize=1.2cm\epsfbox{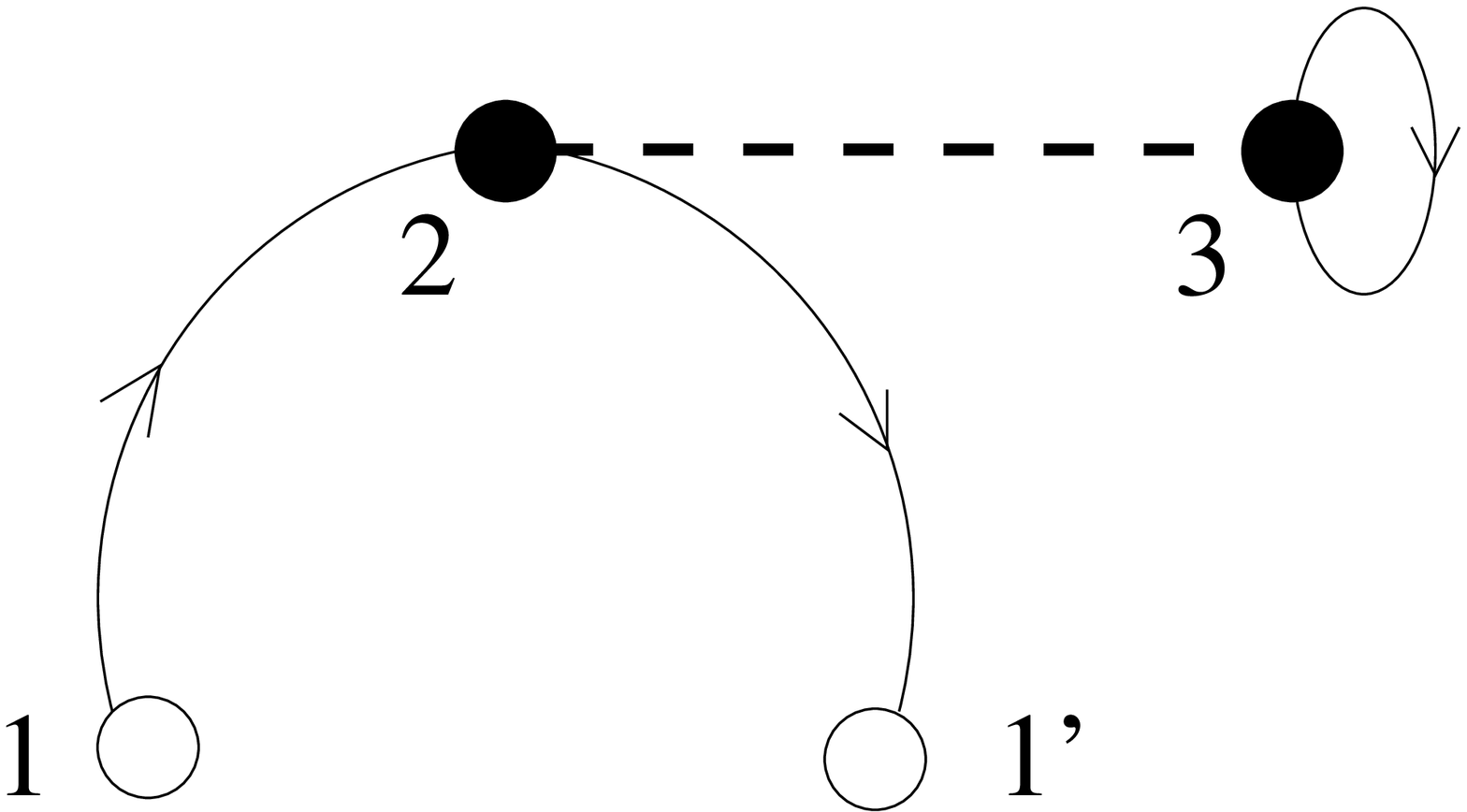}
  \epsfysize=1.2cm\epsfbox{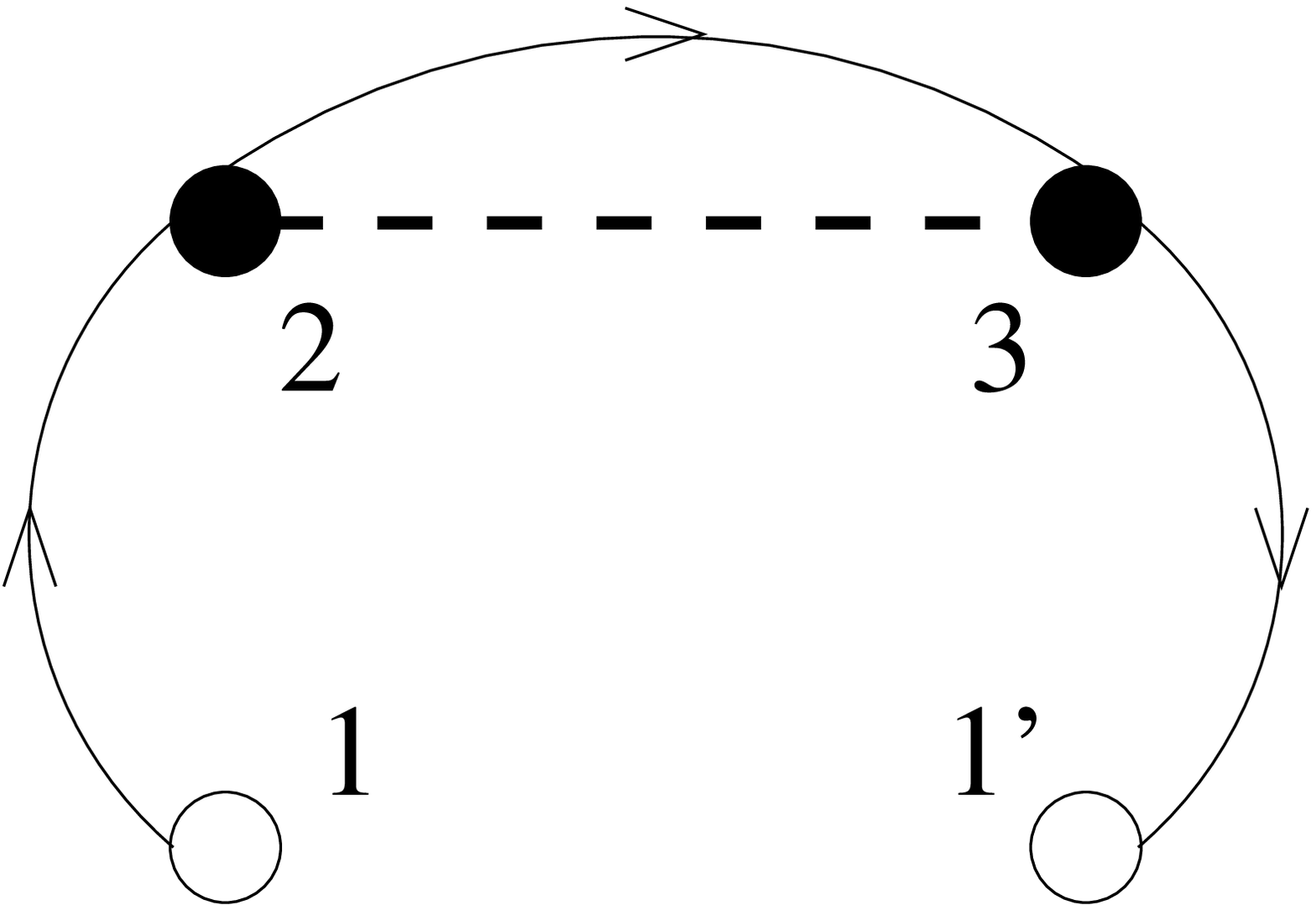}}\normalsize
\centerline{\hspace{-0.1cm}(b)\hspace{3.7cm}(c)}
 \caption{Diagrammatic representation of the one body mixed
density matrix
  $\rho^{(1)}(\Vec{r}_1,\Vec{r}^{\prime}_1)$ in the lowest order of the $\eta$-expansion
  (Eq. (\ref{obdm})). The three sets of diagrams represent the mean field contribution (a),
   and the hole (b)  and spectator (c) direct and exchange
   contributions, respectively. Open dots denote the "active" particles; full
   dots, labeled by an index "i"  stand for an integration over the
   coordinates of particle "i"; an oriented full line, originating
   from  a dot
   and ending in the same dot,  denotes the mean field diagonal OBDM $\rho_o(i)$,
   whereas an oriented full line, joining two different dots,
   represents the non diagonal OBDM density matrix, $\rho_o(i,j)$;
   dashed lines in $(b)$ represent $\hat{\eta}_{11^\prime2}$ and
   those in $(c)$  $\hat{\eta}_{23}$.}\label{Fig0}
    \vskip -0.2cm
\end{figure}
compensated to a large extent by a significantly larger expected
probability of  proton-neutron ($pn$) correlations. In the past
decade  evidence of SRCs has been provided by a new class of
experiments  based upon the scattering of  leptonic and hadronic
probes off nuclei at high momentum transfer ($Q^2 > 1$ $GeV^2$).
The claimed evidence of SRC in these experiments resulted from: i)
the observation  of a scaling behaviour of  the ratios of
inclusive $A(e,e^\prime)X$ cross sections on heavy nuclei to those
on deuteron, for values of the Bjorken variable $1.4 \leq x_B \leq
2$ \cite{ratioAD}, which would indicate that the electron probes
two-nucleon correlations in nuclei similar to the ones in the the
two body system; ii) the observation of approximate scaling of
the ratios of inclusive $A(e,e^\prime)X$ cross sections on heavy
nuclei to those of $^3He$, for values of the Bjorken variable $2
\leq x_B \leq 3$ \cite{ratioA3}, which would represent evidence of
three-nucleon correlations; iii) the observation of $np$ pairs
emitted back-to-back in the process $^{12}C(p,ppn)X$  \cite{EVA}
which provided a direct
\begin{figure}[!hbp]
  \centerline{(a)}
  \centerline{\hspace{-0cm}\epsfysize=0.7cm{\epsfbox{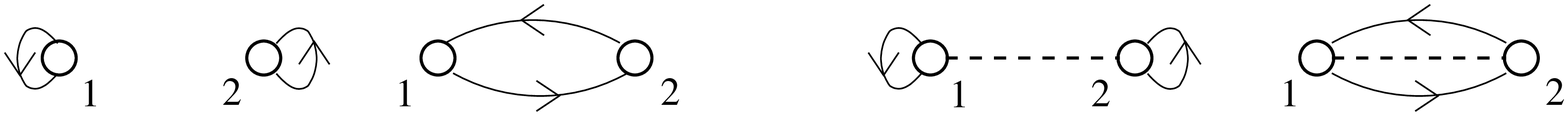}}}
  \centerline{(b)}
  \centerline{\hspace{-0cm}\epsfysize=1.35cm{\epsfbox{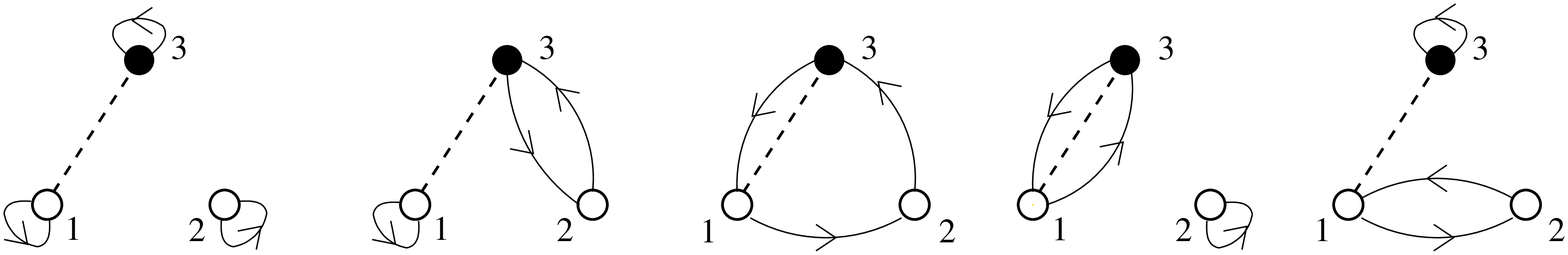}}}
  \centerline{(c)}
  \centerline{\hspace{-0cm}\epsfysize=3.5cm{\epsfbox{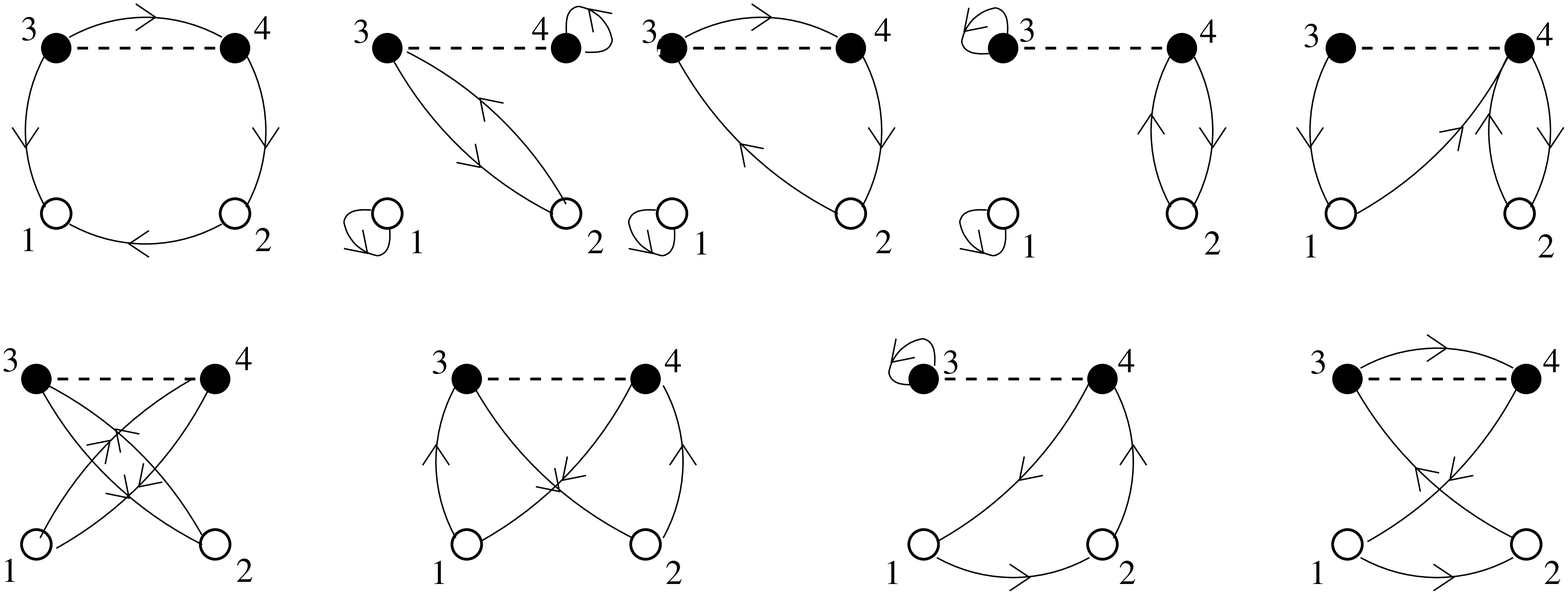}}}
  \vskip -0.2cm
  \caption{The same as in Fig. \ref{Fig0} for the
   \textit{diagonal} two-body density matrix
    (Eq. (\ref{tbdm}) with $\Vec{r}^{\prime}_1=\Vec{r}_1$ and
    $\Vec{r}^{\prime}_2=\Vec{r}_2$). The groups of diagrams
    represent the MF + two- $(a)$, three- $(b)$ and four-body
    $(c)$ contributions, respectively.}\label{Fig1}
  \vskip -0.2cm
\end{figure}
measurement of correlated $np$ pairs,  with a
yield consistent with the $A(e,e^\prime)X$ results; a recent
analysis of these data \cite{pia01} shows that, in agreement with
\begin{figure*}[!htp]
  \centerline{\hspace{-0.3cm}
    \epsfig{width=6.3cm,height=4.5cm,file=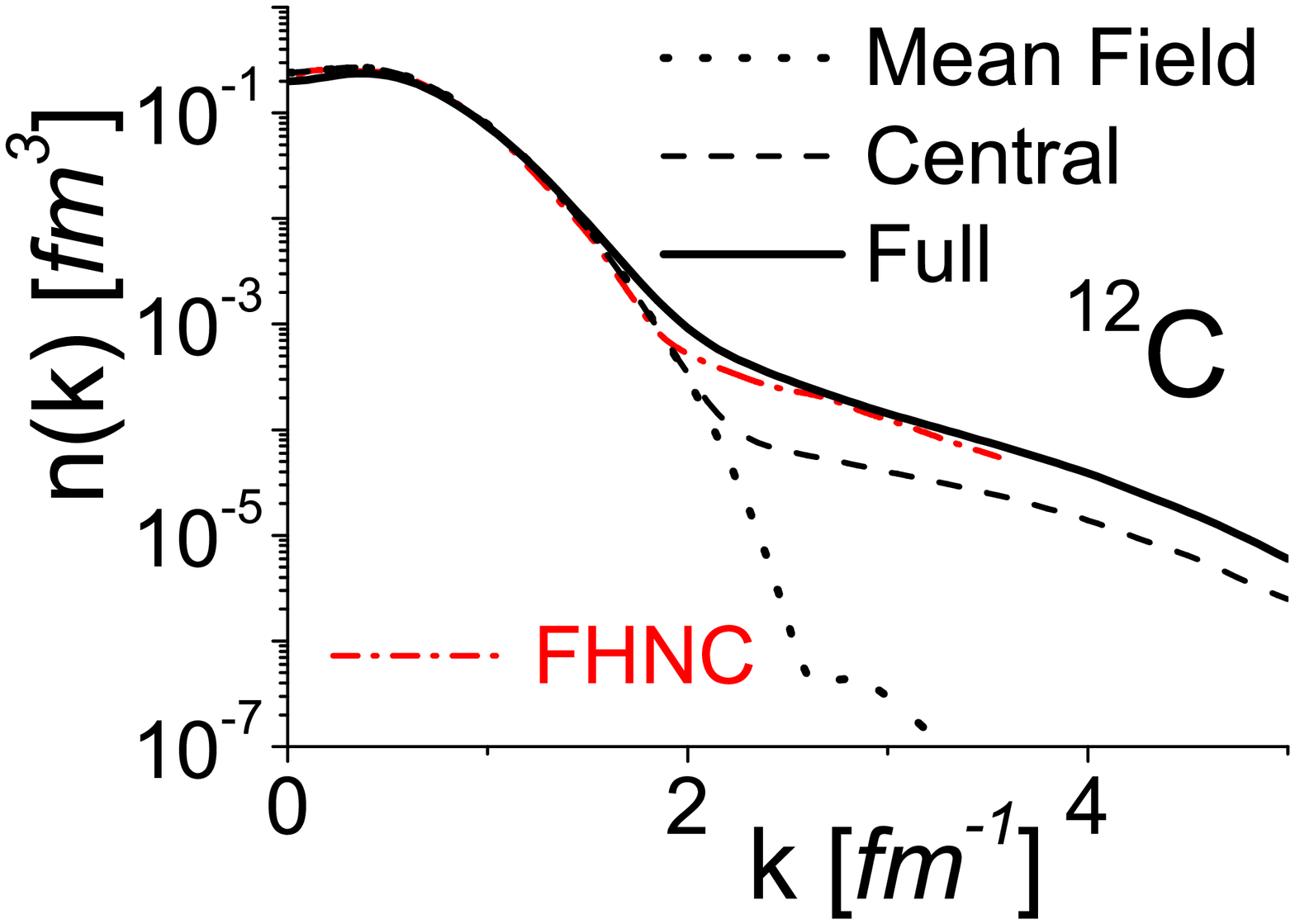}\hspace{-0.8cm}
    \epsfig{width=6.0cm,height=4.5cm,file=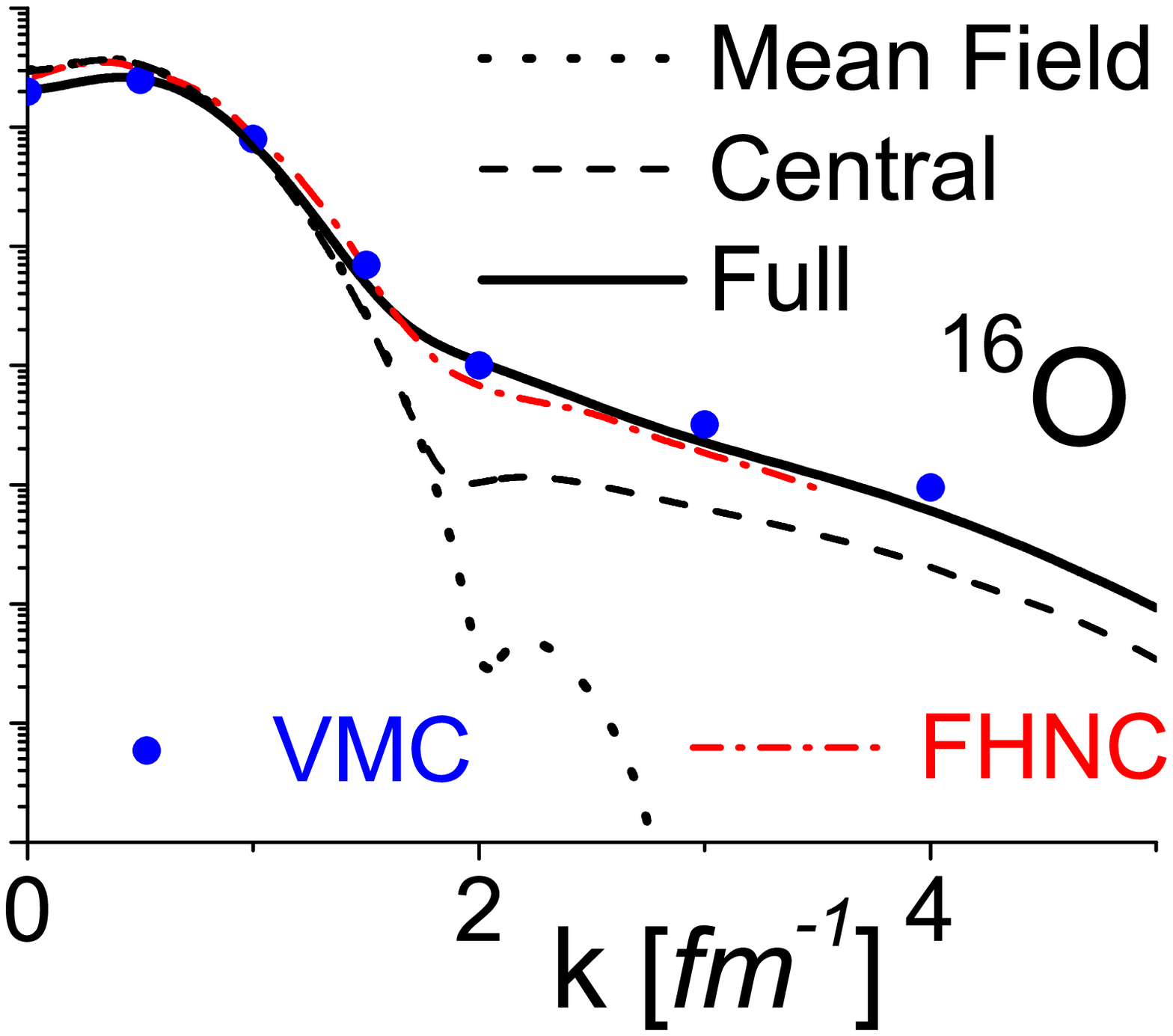}\hspace{-0.6cm}
    \epsfig{width=5.6cm,height=4.5cm,file=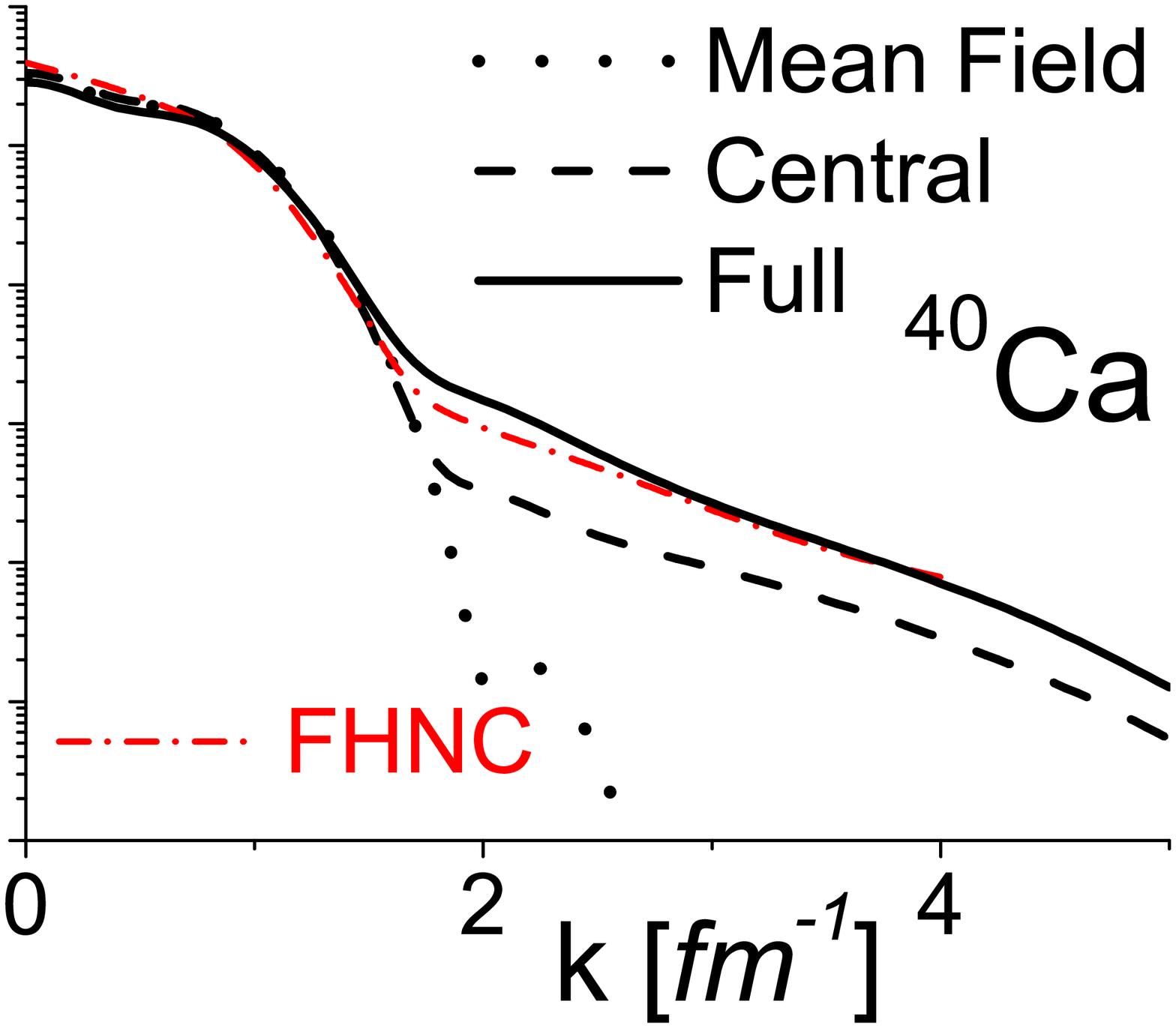}}
  \vskip -0.5cm
  \caption{The one-nucleon momentum distribution
  calculated within the cluster expansion of Ref. \cite{alv1} (dots, dashes and full))
  compared with the
  VMC results of Ref. \cite{pie01} (full dots)  and the FHNC
    results of Ref. \cite{bis01} (dot-dashes).}
    \label{Fig2}
\end{figure*}
\begin{figure*}[!htp]
  \centerline{\hspace{-0.2cm}
    \epsfig{width=6.3cm,height=4.5cm,file=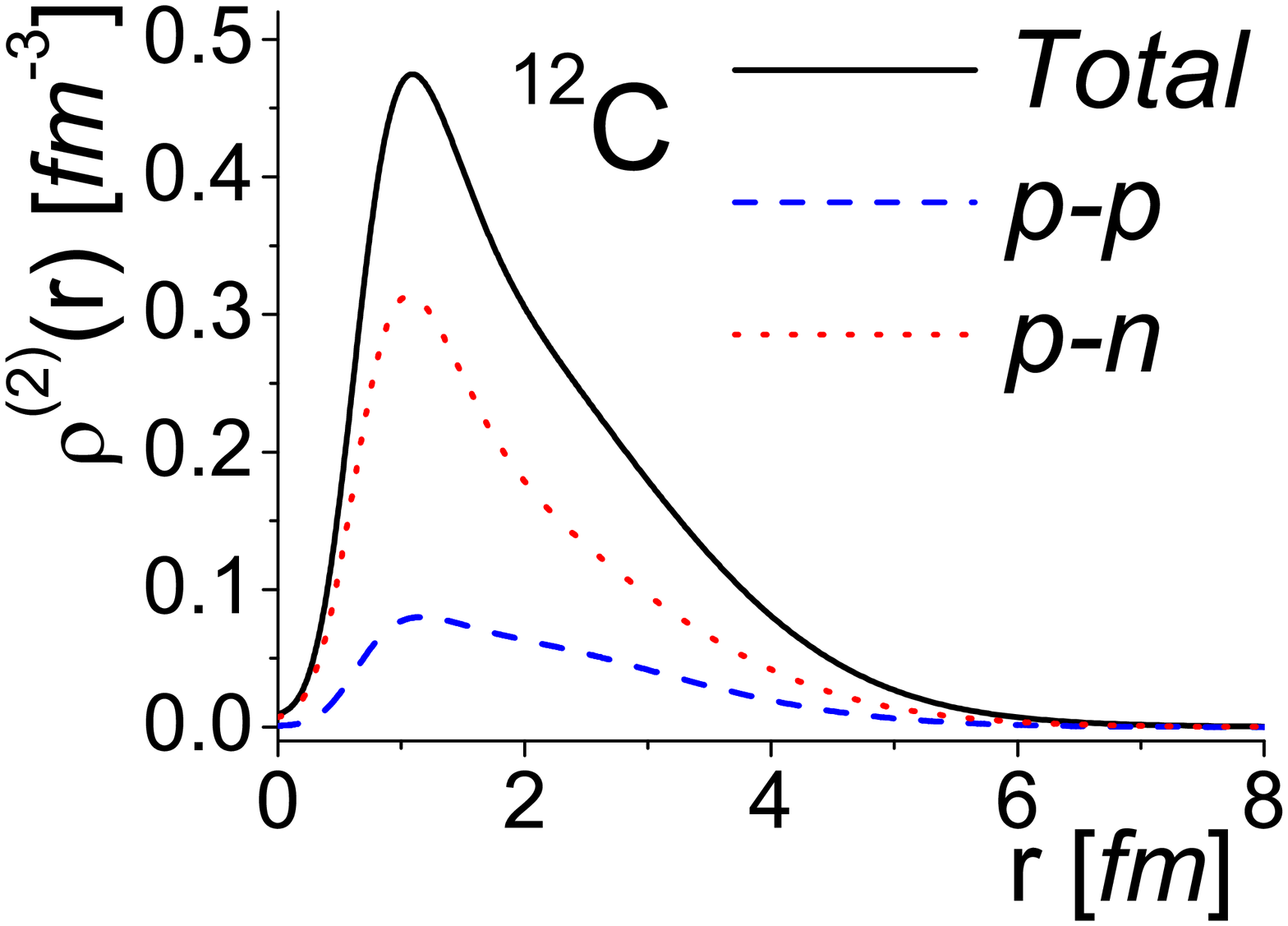}\hspace{-0.6cm}
    \epsfig{width=6.0cm,height=4.5cm,file=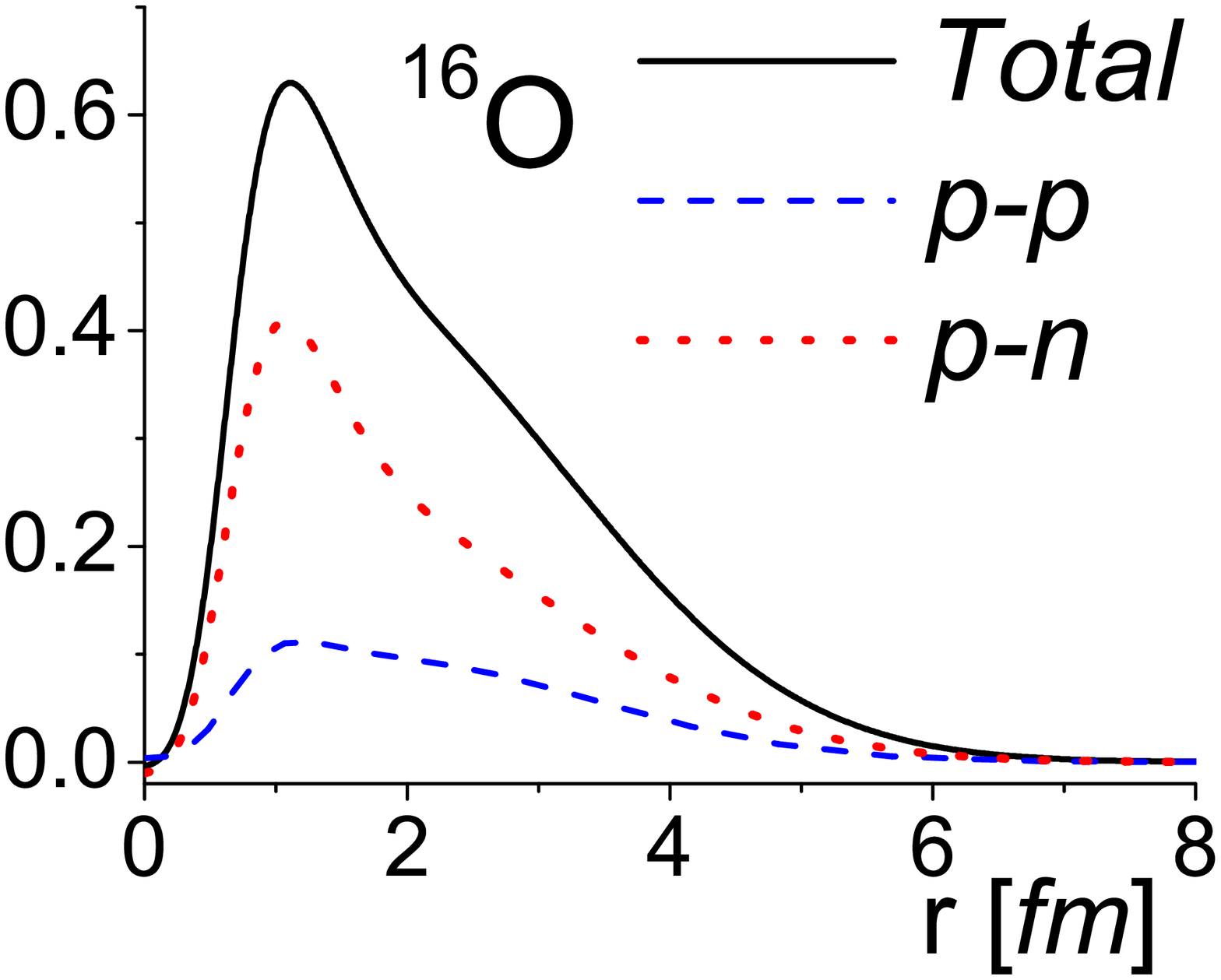}\hspace{-0.6cm}
    \epsfig{width=5.6cm,height=4.5cm,file=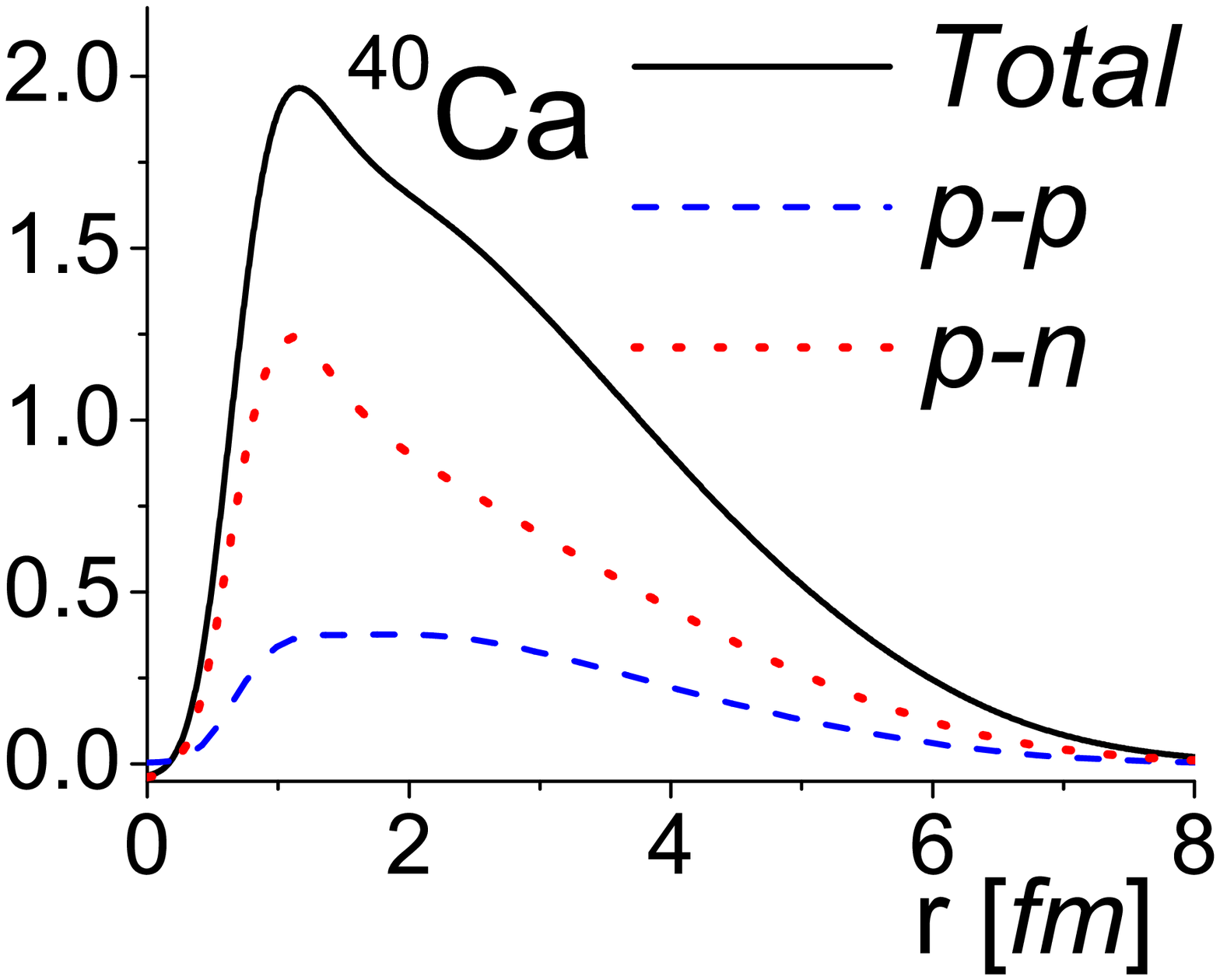}}
  \vskip -0.5cm
  \caption{The relative two-body density distribution  (Eq. (\ref{tbdm})
    with $\Vec{r}^{\prime}_1=\Vec{r}_1$ and $\Vec{r}^{\prime}_2=\Vec{r}_2$)
    integrated over the CM coordinate $\Vec{R}=(\Vec{r}_1+\Vec{r}_2)/2$).
    The proton-neutron (\textit{pn}) and  proton-proton (\textit{pp})
     contributions are also shown. The  various curves are normalized to the
     corresponding numbers of  NN  pairs of the given type.}\label{Fig3}
\end{figure*}
\begin{figure*}[!p]
  \centerline{\hspace{-0.2cm}
    \epsfig{width=6.3cm,height=4.5cm,file=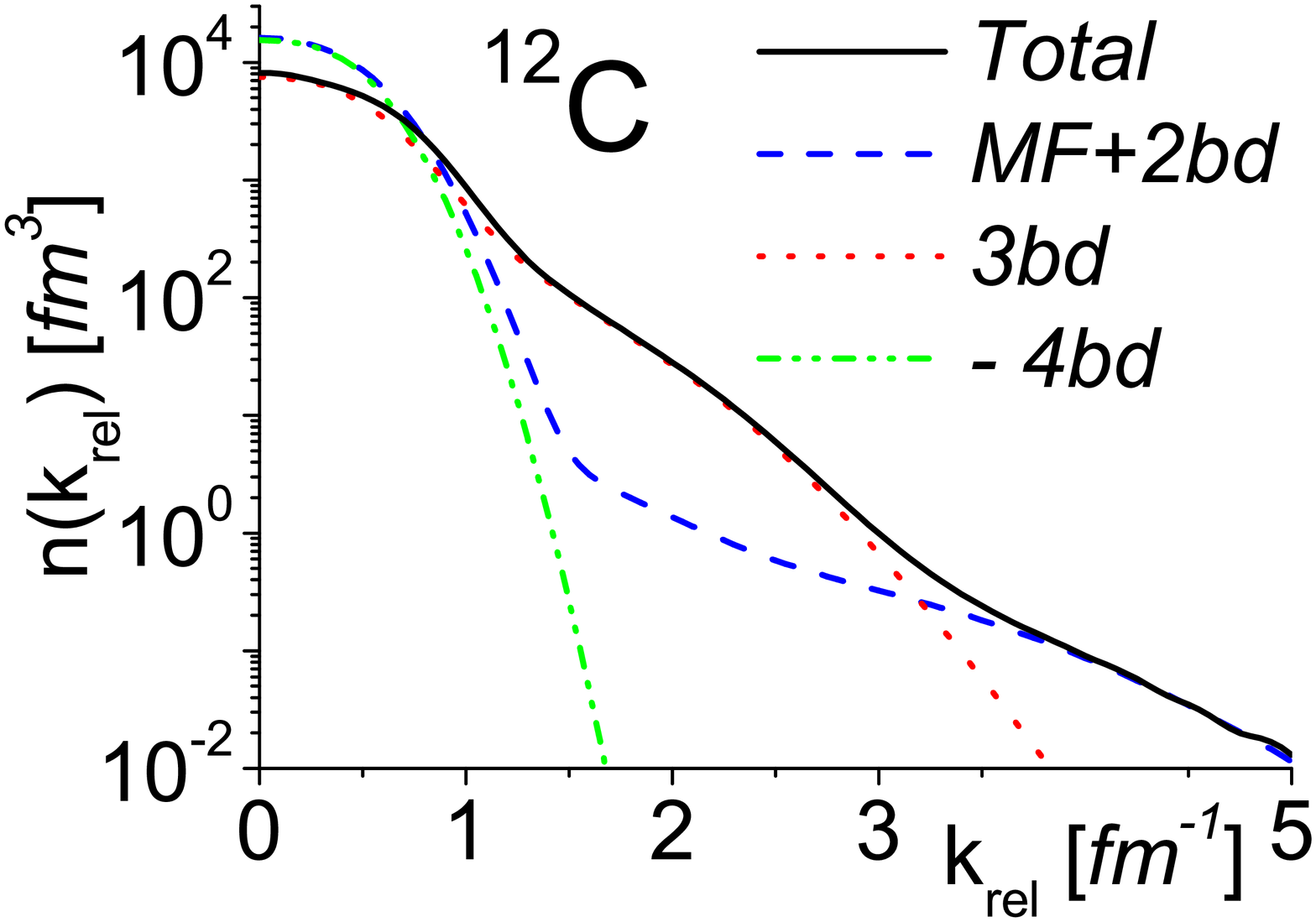}\hspace{-0.6cm}
    \epsfig{width=6.0cm,height=4.5cm,file=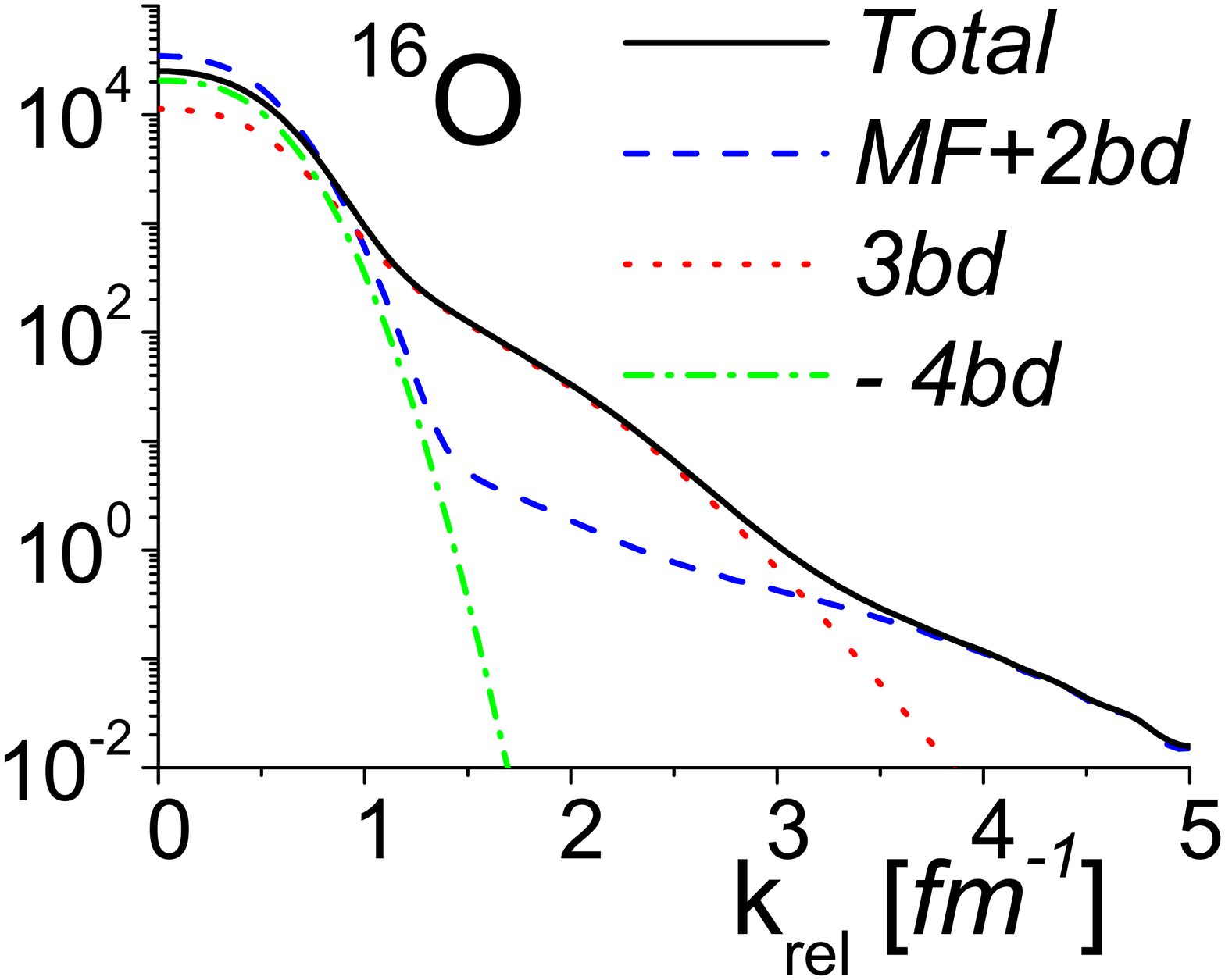}\hspace{-0.6cm}
    \epsfig{width=5.6cm,height=4.5cm,file=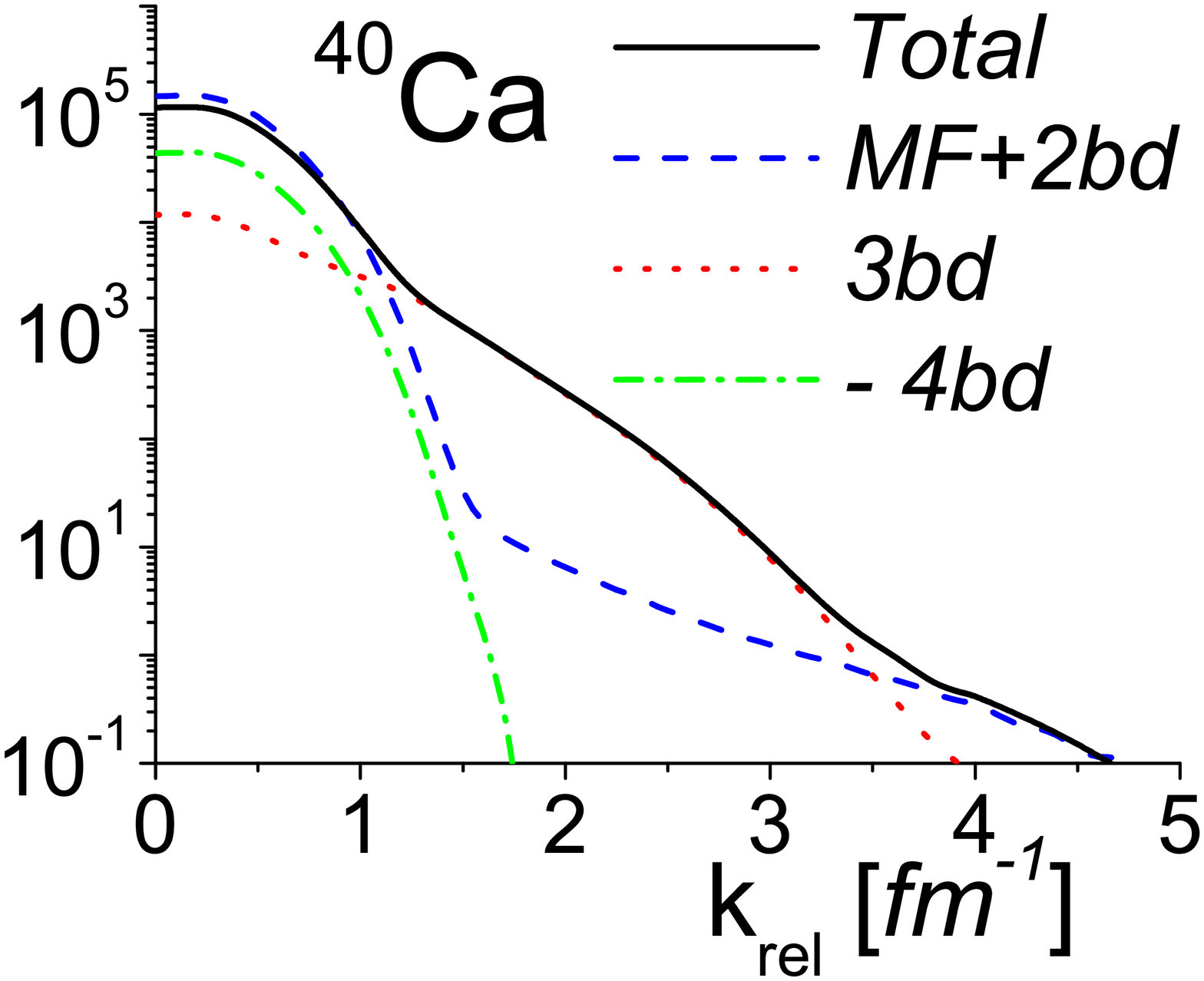}}
  \vskip -0.5cm
  \caption{The relative two-body momentum distribution (Eq. (\ref{2bmomdis}))
   integrated over $\Vec{K}_{CM}$. The various contributions
    corresponding to the diagrams  in Fig.\ref{Fig1} are also shown.}\label{Fig4}
\end{figure*}
\begin{figure*}[!p]
\vskip -0.1cm \centerline{\hspace{-0.3cm}
    \epsfig{width=6.3cm,height=4.5cm,file=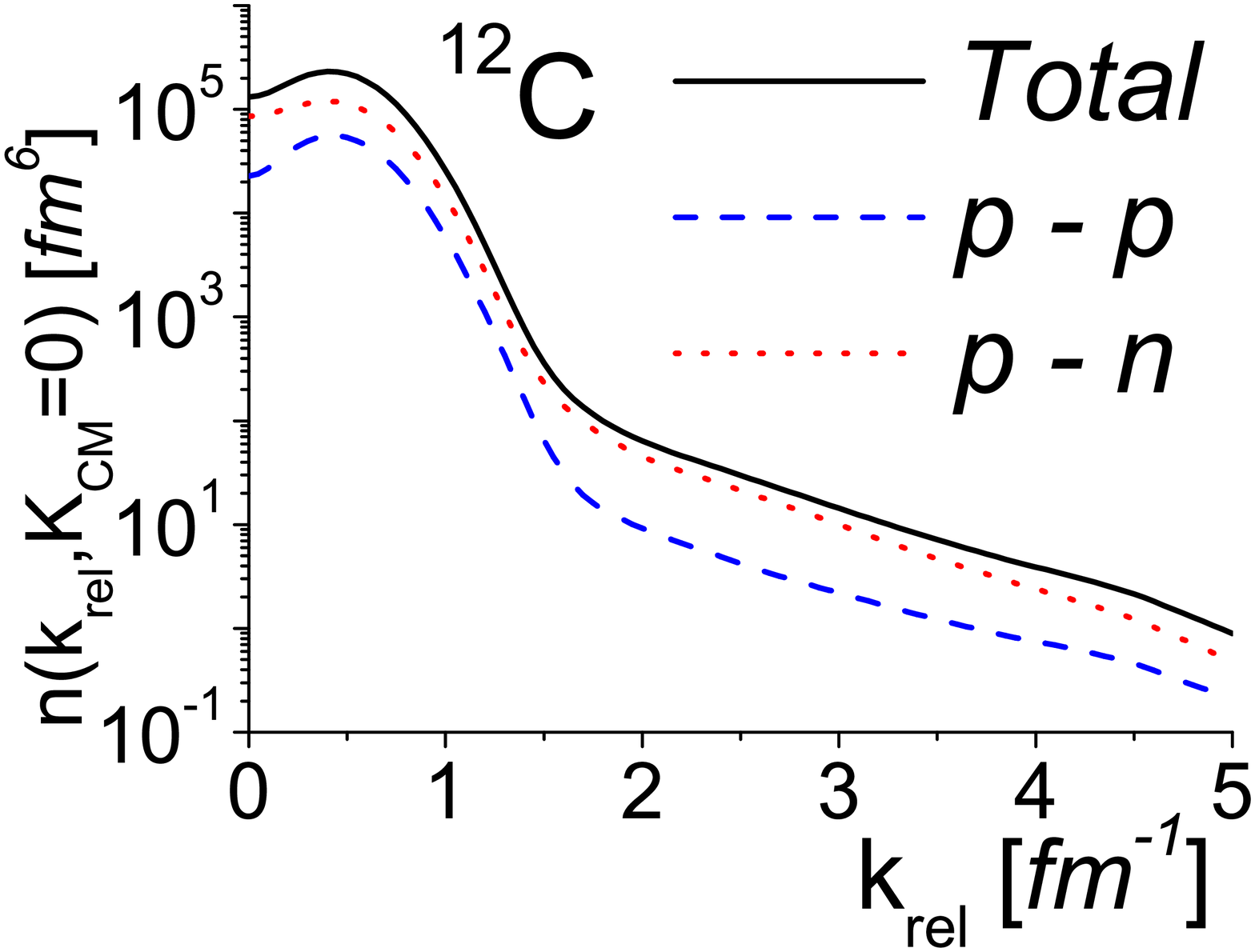}\hspace{-0.6cm}
    \epsfig{width=6.0cm,height=4.5cm,file=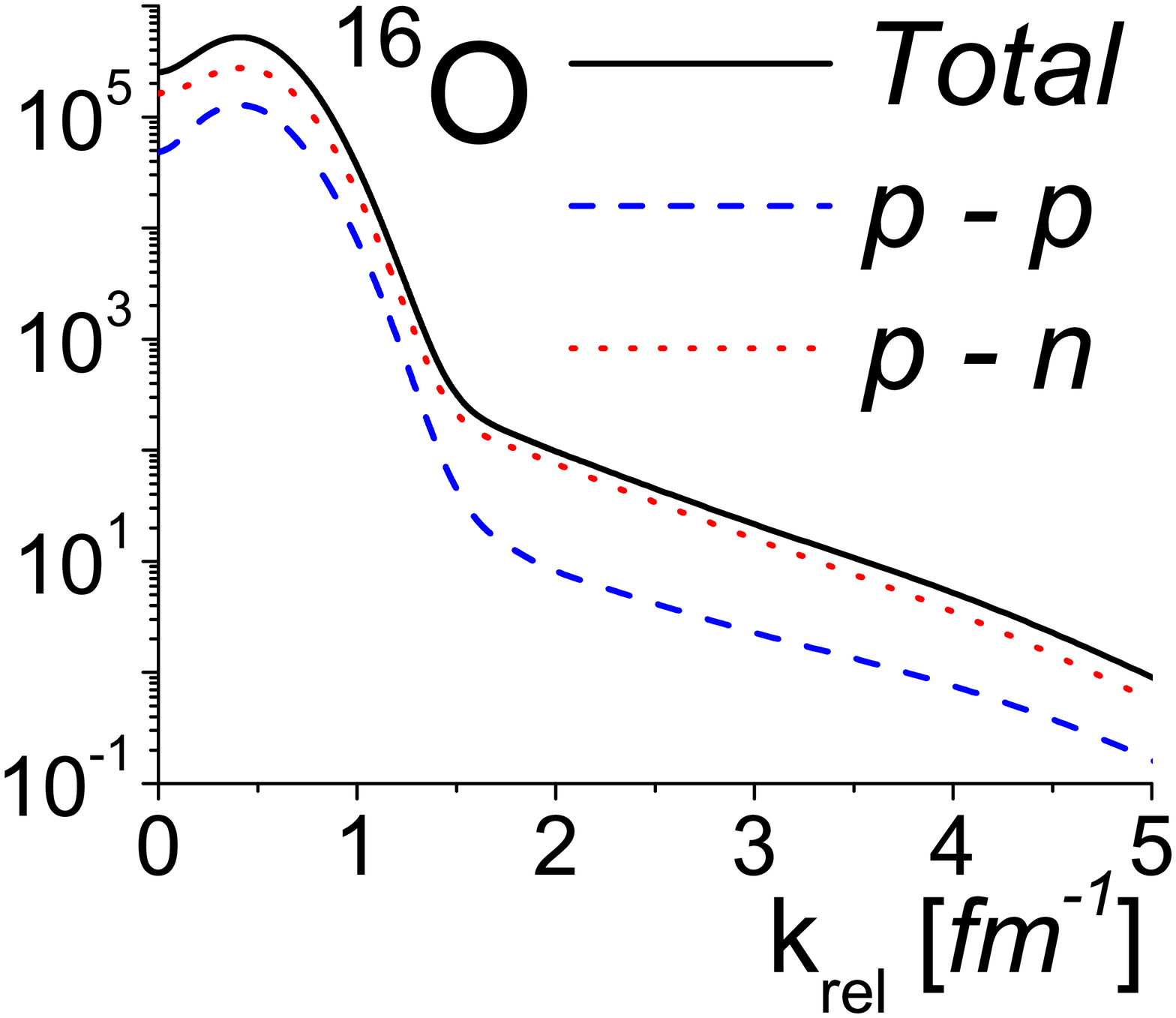}\hspace{-0.6cm}
    \epsfig{width=5.6cm,height=4.5cm,file=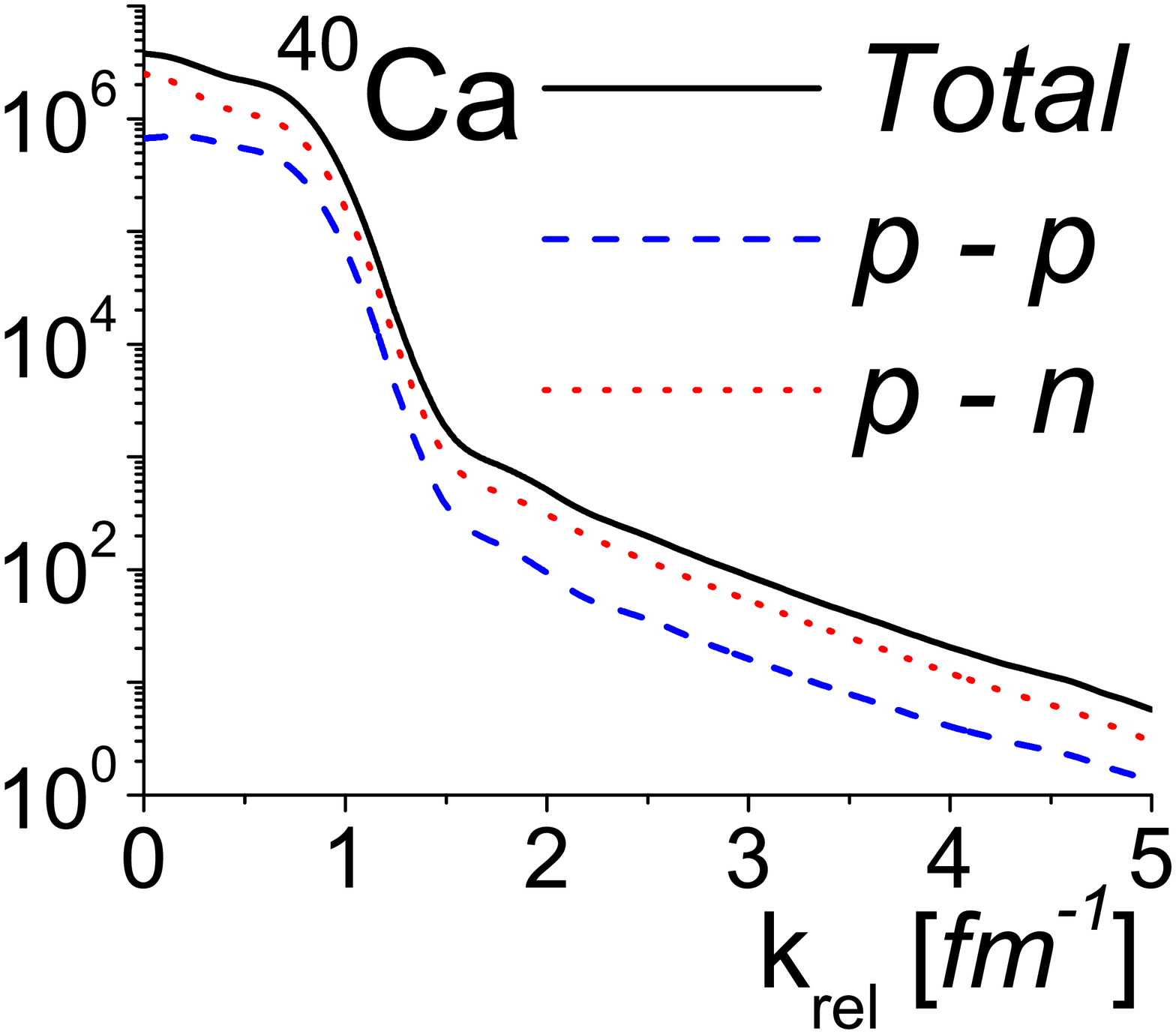}}
  \vskip -0.5cm
\caption{The relative two-body momentum distribution (Eq.
(\ref{2bmomdis})) calculated in correspondence of
  $\Vec{K}_{CM}=0$ (back-to-back nucleons).}
  \label{Fig5}
\end{figure*}
theoretical predictions \cite{mark1,ciosim}, the high  nucleon
momentum tail in nuclei is governed  by  two nucleon SRC dominated
by $np$ correlations (in carbon, over 74\% of protons with momenta
above 275 $MeV/c$ were found to be  members of an $np$ correlated
pair); iv) the direct observation of $pp$ correlated pairs in a
recent JLab experiment \cite{E01-015}, where a simultaneous
measurement of the triple coincidence $^{12}C(e,e^\prime pp)X$ and
the double coincidence $^{12}C(e,e^\prime p)X$ reactions revealed
that the  ratio of $^{12}C(e,e^\prime pp)X$ to $^{12}C(e,e^\prime
p)X$ events, for proton missing momenta above 300 $MeV/c$, is
$9.5\pm 2\%$. Further experimental work on SRC is planned, aimed,
particularly,  at measuring simultaneously  the $^4He(e,e^\prime
pp)X/^4He(e,e^\prime p)X$, $^4He(e,e^\prime pn)X/^4He(e,e^\prime
p)X$, and $^4He(e,e^\prime pn)X/^4He(e,e^\prime pp)X$ ratios for
missing momenta in the range 500-875 $MeV/c$, in order to
investigate the  hard core region \cite{pia02}. The experimental
results cited above, in particular  the   large ratio of $pn$ to
$pp$ SRC, observed by the EVA/BNL \cite{EVA} and E01-015 \cite{E01-015}
experiments, call for a solid theoretical validation. In Ref. \cite{rocco}
the  role of the tensor force in producing a substantial difference
between $pn$ and $pp$ two-nucleon momentum distributions in few-body
systems and light nuclei ($A<8$) has been analyzed using {\it
status-of-the-art} realistic nuclear wave functions obtained
within the Variational Monte Carlo (VMC) approach \cite{VMC}. In
this paper  the results of calculations \cite{alv3} of the effect
of the tensor force on  the one- and two-body momentum
distributions of medium weight nuclei ($12\leq A \leq 40$),
obtained within a linked cluster expansion and realistic
interactions \cite{alv1,alv2}, will be presented. In what follows,
the cluster expansion method for the one- and two-body
densities is outlined; the one- and two-body momentum
distributions are illustrated and, eventually,  the
results of calculations are presented. Within the linked cluster expansion of Ref.
\cite{alv1,alv2}, the expression for
the non diagonal one body density matrix (OBDM) resulting from a correlated
ground state wave function of the form $\psi_o = \phi_o \prod \hat{f}(ij)$ is,
in lowest order of $\hat{\eta}_{ij} = \hat{f}^2_{ij}-1$:
\beq\label{obdm}
\rho^{(1)}(\Vec{r}_1,\Vec{r}_1^\prime)\,=\,\rho^{(1)}_{SM}(\Vec{r}_1,\Vec{r}_1^\prime)
        \,+\,\rho^{(1)}_{H}(\Vec{r}_1,\Vec{r}_1^\prime)
        \,+\,\rho^{(1)}_{S}(\Vec{r}_1,\Vec{r}_1^\prime)\,,
\eeq where $\rho_{SM}(\Vec{r},\Vec{r}^\prime) = \sum_\alpha
\varphi^\star_\alpha(\Vec{r}) \varphi_\alpha(\Vec{r}^\prime)$ is
the mean field (MF)   density, $\varphi_{\alpha}(\Vec{r})$ the MF
wave function, and $\rho^{(1)}_{H}$ and $\rho^{(1)}_{S}$ the hole
(H) and spectator (S) correlation contributions,  given
respectively by (see Ref. \cite{alv1})
\begin{widetext}
\beqy
\rho^{(1)}_{H}(\Vec{r}_1,\Vec{r}_1^\prime)&=&\int\,d\Vec{r}_2\,\Big[H_D(\Vec{r}_{12},\Vec{r}_{1^\prime2})
  \,\rho_o(\Vec{r}_1,\Vec{r}_1^\prime)
  \,\rho_o(\Vec{r}_2)\,-\,H_E(\Vec{r}_{12},\Vec{r}_{1^\prime2})\,\rho_o(\Vec{r}_1,\Vec{r}_2)
  \,\rho_o(\Vec{r}_2,\Vec{r}_1^\prime)\Big]\nonumber\\
\rho^{(1)}_{S}(\Vec{r}_1,\Vec{r}_1^\prime)&=&-\int d\Vec{r}_2d\Vec{r}_3
\rho_o(\Vec{r}_1,\Vec{r}_2)\Big[H_D(\Vec{r}_{23})\,
  \rho_o(\Vec{r}_2,\Vec{r}_1^\prime)\rho_o(\Vec{r}_3)\,
  -\,H_E(\Vec{r}_{23})\,\rho_o(\Vec{r}_2,\Vec{r}_3)\rho_o(\Vec{r}_3,
  \Vec{r}_1^\prime)\Big]\,.\label{2}
\eeqy
\end{widetext}
Here  the direct (D) and exchange (E) functions $H_{D,E}$ are the
expectation values of the correlation operators with respect to
the spin-isospin functions, with the dependence upon the
coordinates originating from the spatial part of the correlation
functions and from the tensor operator (the explicit expressions
are given  in Ref. \cite{alv1}). The non diagonal two-body density
matrix (TBDM) can be written as follows:
\begin{widetext}
\beqy\label{tbdm}
\rho^{(2)}(\Vec{r}_1,\Vec{r}_2;\Vec{r}^\prime_1,\Vec{r}^\prime_2)&=&
\rho^{(2)}_{SM}(\Vec{r}_1,\Vec{r}_2;\Vec{r}^\prime_1,\Vec{r}^\prime_2)
\,+\,\rho^{(2)}_{2b}(\Vec{r}_1,\Vec{r}_2;\Vec{r}^\prime_1,\Vec{r}^\prime_2)
\,+\,\rho^{(2)}_{3b}(\Vec{r}_1,\Vec{r}_2;\Vec{r}^\prime_1,\Vec{r}^\prime_2)
\,+\,\rho^{(2)}_{4b}(\Vec{r}_1,\Vec{r}_2;\Vec{r}^\prime_1,\Vec{r}^\prime_2)
\eeqy
\end{widetext}
where $\rho^{(2)}_{SM}$ represents the mean field contribution and
the other terms the correlations contributions, with the
subscripts $2b$, $3b$ and $4b$ denoting  the number of "bodies"
(nucleons) involved in the given contribution. The various terms
appearing in Eq. (\ref{obdm}) are diagrammatically represented  in
Fig. \ref{Fig0}, whereas the diagrams corresponding to  the
diagonal TBDM  (Eq. (\ref{tbdm}) with $\Vec{r}^\prime_1=\Vec{r}_1$
and $\Vec{r}^\prime_2=\Vec{r}_2$) are presented in Fig.
\ref{Fig1}.  The explicit expressions of the diagonal and
non-diagonal TBDM are given in Ref. \cite{alv3} in terms
correlation functions and mean field non diagonal one-body density
matrices. In case of central correlations (and harmonic oscillator
(HO)  mean field wave functions) analytic expressions of the
diagonal TBDM are given in Ref. \cite{dim01}; when non central
correlations are considered, as in the present case, no analytic
expression can be given even using HO wave functions. The
one-nucleon momentum distribution (NMD) is the Fourier transform
of the OBDM (Eq. (\ref{2})), {\it viz} \beq \label{momdis1}
n(\Vec{k})\,=\,\frac{1}{(2\pi)^3}\,\int
d\Vec{r}_1d\Vec{r}^\prime_1\,
e^{-i\,\Vec{k}\cdot\left(\Vec{r}_1\,-\,\Vec{r}^\prime_1\right)}\,
\rho^{(1)}(\Vec{r}_1,\Vec{r}^\prime_1)\,, \eeq whereas  the
Fourier transform of the TBDM (Eq. (\ref{tbdm})), {\it i.e.}
\begin{widetext}
\beq\label{momdis2}
n(\Vec{k}_1,\Vec{k}_2)\,=\,\frac{1}{(2\pi)^6}\int
d\Vec{r}_1d\Vec{r}_2d\Vec{r}^\prime_1d\Vec{r}^\prime_2
e^{i\Vec{k}_1\cdot\left(\Vec{r}_1-\Vec{r}^\prime_1\right)}
\,e^{i\Vec{k}_2\cdot\left(\Vec{r}_2-\Vec{r}^\prime_2\right)}
\rho^{(2)}\left(\Vec{r}_1,\Vec{r}_2;\Vec{r}^\prime_1,\Vec{r}^\prime_2\right)
\eeq \end{widetext}
represents the two-body momentum distribution
(2NMD), which can be written in terms of relative
($\Vec{k}_{rel}$) and Center-of-Mass ($\Vec{K}_{CM}$) momenta as
follows
\begin{widetext}
\beq
\label{2bmomdis} n(\Vec{k}_{rel},\Vec{K}_{CM})\,=\,
\,\frac{1}{(2\pi)^6}\int
d\Vec{r}\,d\Vec{R}\,d\Vec{r}^\prime\,d\Vec{R}^\prime\,
e^{i\,\Vec{K}_{CM}\cdot\left(\Vec{R}-\Vec{R}^\prime\right)}\,
e^{i\,\Vec{k}_{rel}\cdot\left(\Vec{r}-\Vec{r}^\prime\right)}\,
\rho^{(2)}(\Vec{r},\Vec{R};\Vec{r}^\prime,\Vec{R}^\prime)\,, \eeq
\end{widetext}
where $\Vec{r}=\Vec{r}_1-\Vec{r}_2$,
$\Vec{r}^\prime=\Vec{r}^\prime_1-\Vec{r}^\prime_2$,
$\Vec{R}=(\Vec{r}_1+\Vec{r}_2)/2$ and
$\Vec{R}^\prime=(\Vec{r}^\prime_1+\Vec{r}^\prime_2)/2$. The
relative,  $n_{rel}(\Vec{k}_{rel})$,  and CM,
$n_{CM}(\Vec{K}_{CM})$, momentum distributions can be obtained
from Eq. (\ref{2bmomdis}) integrating over $\Vec{K}_{CM}$ and
$\Vec{k}_{rel}$, respectively. In the numerical calculations we
have considered the case of $A=12, 16$ and $40$. The ingredients
we need for  the calculations are the single particle MF wave
functions and the correlation functions, appearing in  Eqs.
(\ref{obdm}) and (\ref{tbdm}). The correlation functions we have
used include central, spin, isospin and tensor correlations and
correspond to the $V8^\prime$ realistic
\begin{figure}[!hbp]
  \vskip -0.5cm
  \centerline{\epsfxsize=6.5cm\epsfbox{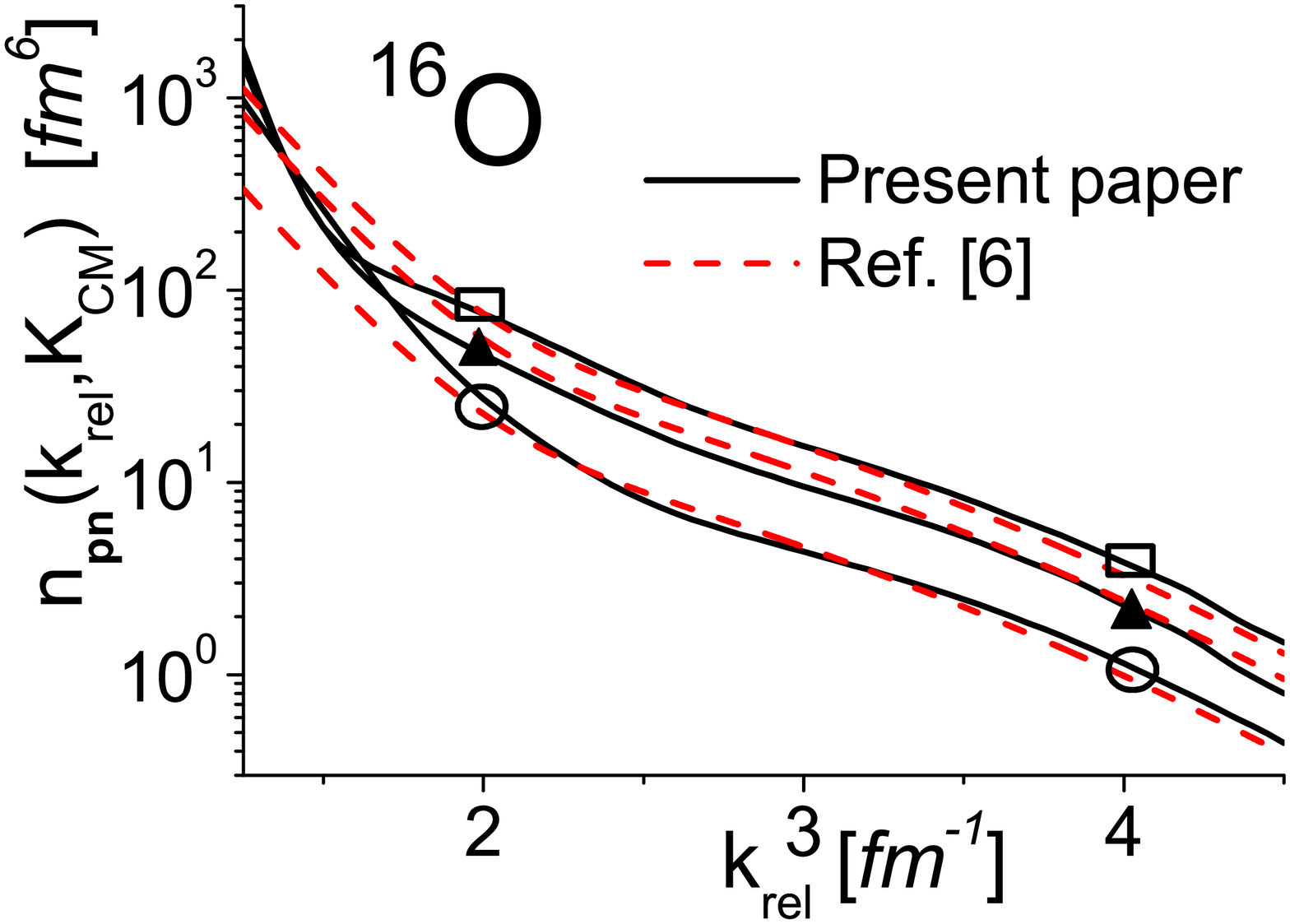}}
  \vskip -0.5cm
  \centerline{\epsfxsize=6.2cm\epsfbox{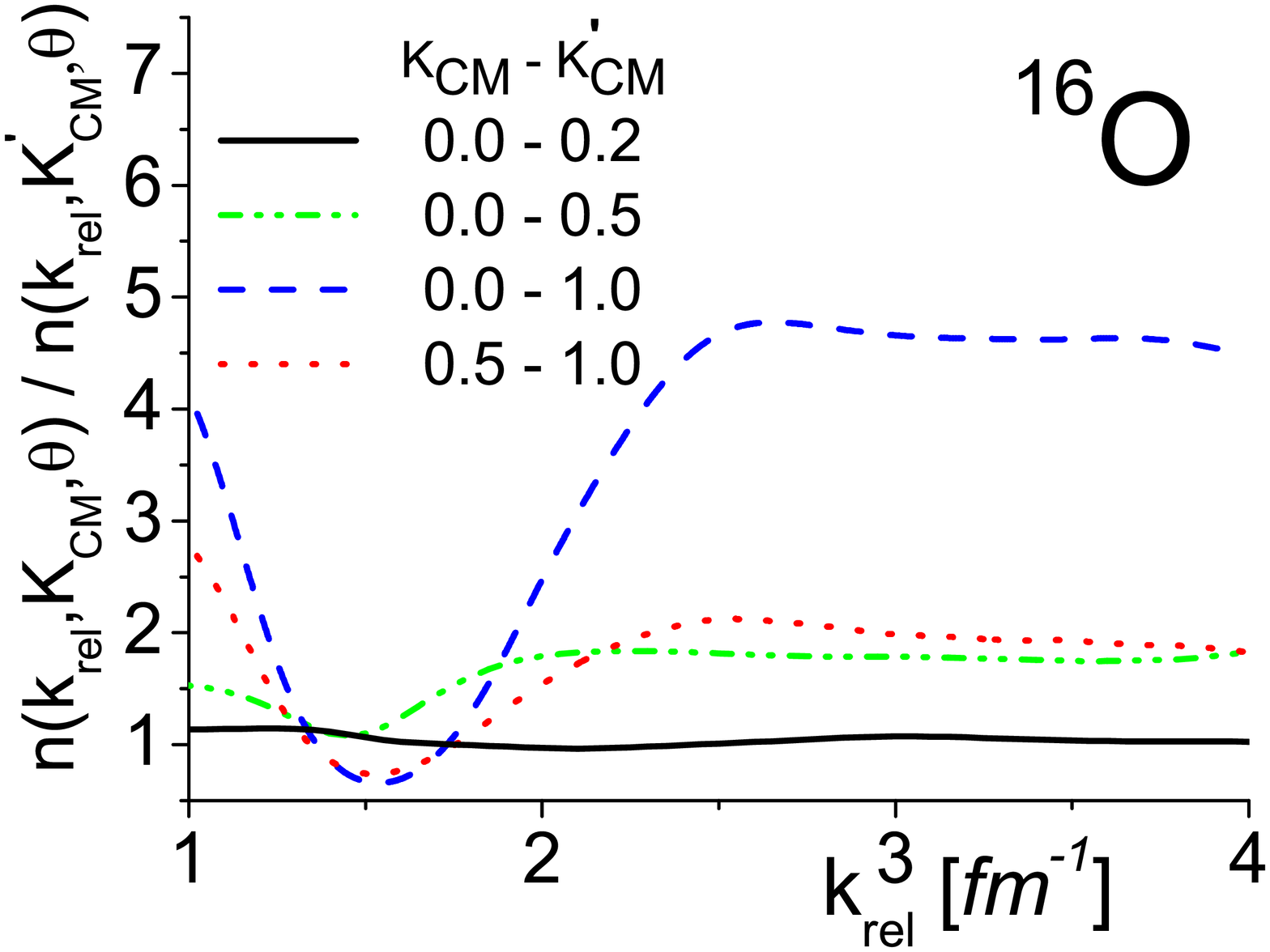}}
  \vskip -0.5cm
\caption{Check of the factorization assumption of the 2NMD.  \textit{Left}: the exact 2NMD
  of the present paper  compared with the factorized form of Ref. \cite{ciosim}
  $n(\Vec{k}_{rel},\Vec{K}_{CM}) = C_A\, n_D(k_{rel})\, n_{CS}(K_{CM})$ (see text); the values
  of $K_{CM}$ corresponding to the various curves are $K_{CM} = \,0$ $fm^{-1}\, (squares),\,
  \, 0.5$ $fm^{-1} (full\,\, triangles),\, 1$ $fm^{-1}\, (open\,\, dots)$. \textit{Right}: the
  ratio $n(k_{rel},K_{CM},\theta)/n(k_{rel},K^\prime_{CM},\theta)$ for various values of $K_{CM}$
  and  $K^\prime_{CM}$ and a fixed value of  $\theta=90^o$ (see text).}
\label{Fig6}
\end{figure}
interaction \cite{vuotto}; the MF  wave functions, whose
parameters were fixed  by minimizing the ground state energy (see
\cite{alv1}),  have been chosen both in the harmonic oscillator and
Woods-Saxon forms. The momentum distributions of $^{12}C$, $^{16}O$ and
$^{40}Ca$ are shown in Fig. \ref{Fig2}, where they are compared with the
results of higher order calculations, e.g. the Variational Monte Carlo
and Fermi Hyper Netted Chain ones.The results for the diagonal TBDM
are  shown in Fig. \ref{Fig3}, and the $pn$, $pp$ and total
two-body momentum distributions are shown in Figs. \ref{Fig4} and
Fig. \ref{Fig5}; eventually, in Fig. \ref{Fig6} the extent to
which the 2NMD factorizes into the relative and CM momentum
distributions is illustrated. The factorization of the 2NMD has
been used in Ref. \cite{ciosim} to obtain the one nucleon spectral
function of complex nuclei at high values of momenta and removal
energies. We have checked such a factorization assumption by: i)
comparing the exact 2NMD with the factorized form used in
\cite{ciosim}, {\it viz} $n(\Vec{k}_{rel},\Vec{K}_{CM}) = C_A\,
n_D(k_{rel})\, n_{CS}(K_{CM})$, where $n_D(k_{rel})$ is the
deuteron momentum distribution and $n_{CS}(K_{CM})$ the CM
distribution given in \cite{ciosim};  ii) by analyzing the  ratio
$n(k_{rel},K_{CM}, \theta)/n(k_{rel},K^\prime_{CM},\theta)$ for
various values of $K_{CM}$, $K^\prime_{CM}$, and fixed value of
$\theta \equiv \theta_{\widehat{{\bf k}{\bf K}}}=90^o$.

The main results we have obtained can be summarized as follows:
\begin{itemize}
\item the high momentum part of the nucleon momentum distributions clearly exhibits the effects
of tensor
correlations, which can be very reliably described within the the lowest order cluster expansion
we have developed (cf. Fig. \ref{Fig2}) (the effects of the higher order terms is negligibly small
(see Ref.\cite{alv1});
\item in agreement with the results for light nuclei \cite{rocco}, at relative momentum
$k_{rel} \geq
1.5 fm^{-1}$ the momentum distribution
of $pn$ pairs is much larger than that of $pp$ pairs (cf. Fig. \ref{Fig4}), whereas
at small values of $k_{rel}$ the ratios of $pn$ and $pp$ momentum distributions is similar
to the ratio of the number of $pn$ to $pp$ pairs, which is $12/5$, $16/7$ and $40/19$ for
$^{12}C$,  $^{16}O$ and $^{40}Ca$, respectively;
\item our parameter-free many-body approach  confirms the  validity of the factorization approximation
of the 2NMD at high values of the relative momentum and low values of the CM momentum
(\textit{cf.} Fig. \ref{Fig6}).
\end{itemize}


\begin{thebibliography}{99}
%
\bibitem{ratioAD} L. L. Frankfurt, M. I. Strikman, D. B. Day and M. Sargsian,
\textit{Phys. Rev.} \textbf{C48} (1993) 2451;
%
\bibitem{ratioA3} K. S. Egiyan {\it et al} (CLAS) \textit{Phys. Rev. Lett.} \textbf{96}
(2006) 082501
%
\bibitem{EVA}  A. Tang {\it et al}, \textit{Phys. Rev. Lett.} \textbf{90} (2003) 042301
%
\bibitem{pia01} E. Piazetsky, M. Sargsian, L. Frankfurt, M. Strikman and J, W. Watson
  \textit{Phys. Rev. Lett.} \textbf{97} (2006) 162504
%
\bibitem{mark1}  L. Frankfurt M. Strikman \textit{Phys. Rep.} \textbf{76} (1981) 214
%
\bibitem{ciosim} C. Ciofi degli Atti and S. Simula,
  \textit{Phys. Rev.} \textbf{C53} (1996) 1689
%
\bibitem{E01-015}  R. Shneor {\it et al},
  \textit{Phys. Rev. Lett.} \textbf{99} (2007) 072501
%
\bibitem{pia02} E. Piazetsky and D. Higinbotham, JLab experiment E07-006 (2007)
\bibitem{rocco}  R. Schiavilla, R.B. Wiringa, S. C. Pieper and J. Carlson
  \textit{Phys. Rev. Lett.} \textbf{98} (2007) 132501
%
\bibitem{VMC}  R.B. Wiringa and  S.C. Pieper
      \textit{Phys. Rev. Lett.} \textbf{89} (2002) 182501;
%
\bibitem{alv3} M. Alvioli, C. Ciofi degli Atti and  H. Morita, {\it to appear};
%
\bibitem{alv1} M. Alvioli, C. Ciofi degli Atti and H. Morita,
  \textit{Phys. Rev.} \textbf{C72} (2005) 054310;
  \textit{Fizica} \textbf{B13} (2004) 585;
%
  \bibitem{alv2}
  M. Alvioli, C. Ciofi degli Atti and H. Morita,
  in \textit{Science and Supercomputing in Europe 2005}; \textit{Alberigo, Erbacci and
    Garofalo Editors}, Bologna (2005)
%
\bibitem{dim01} S.S. Dimitrova, D.N. Kadrev, A.N. Antonov and M.V. Stoitsov,
  \textit{Eur. Phys. J.} \textbf{A7} (2000) 335
%
\bibitem{vuotto} R. B. Wiringa, V. G. J. Stocks and R. Schiavilla, \textit{Phys. Rev.} \textbf{C51} (1995)
 38;
%
\bibitem{pie01} S.C. Pieper, R.B. Wiringa and V.R. Pandharipande,
      \textit{Phys. Rev.} \textbf{C46}, 1741 (2000).
%
\bibitem{bis01} C. Bisconti, F. Arias De Saavedra and G. C\'o,
  \texttt{nucl-th/0702061}

\end{thebibliography}
\end{document}